\documentclass[%
 reprint,
superscriptaddress,
 amsmath,amssymb,
 aps,
prb,
]{revtex4-1}

\usepackage{graphicx}
\usepackage{epstopdf}
\usepackage{dcolumn}
\usepackage{bm}
\usepackage{gensymb}
\usepackage{upgreek}
\usepackage{hyperref}

\begin{document}

\preprint{APS/123-QED}

\title{Intrinsic Insulating Ground State in Transition Metal Dichalcogenide TiSe$_2$}

\author{Daniel~J.~Campbell}
\affiliation{Center for Nanophysics and Advanced Materials, Department of Physics, University of Maryland, College Park, Maryland 20742, USA}
\author{Chris~Eckberg}
\affiliation{Center for Nanophysics and Advanced Materials, Department of Physics, University of Maryland, College Park, Maryland 20742, USA}
\author{Peter~Y.~Zavalij}
\affiliation{Department of Chemistry, University of Maryland, College Park, Maryland 20742, USA}
\author{Hsiang-Hsi~Kung}
\affiliation{Quantum Matter Institute, University of British Columbia, Vancouver, BC V6T 1Z4, Canada}
\affiliation{Department of Physics \& Astronomy, University of British Columbia, Vancouver, BC V6T 1Z1, Canada}
\author{Elia~Razzoli}
\affiliation{Quantum Matter Institute, University of British Columbia, Vancouver, BC V6T 1Z4, Canada}
\affiliation{Department of Physics \& Astronomy, University of British Columbia, Vancouver, BC V6T 1Z1, Canada}
\author{Matteo~Michiardi}
\affiliation{Quantum Matter Institute, University of British Columbia, Vancouver, BC V6T 1Z4, Canada}
\affiliation{Department of Physics \& Astronomy, University of British Columbia, Vancouver, BC V6T 1Z1, Canada}
\affiliation{Max Planck Institute for Chemical Physics of Solids, 01187 Dresden, Germany}
\author{Chris~Jozwiak}
\affiliation{Advanced Light Source, Lawrence Berkeley National Laboratory, Berkeley, CA 94720, USA}
\author{Aaron~Bostwick}
\affiliation{Advanced Light Source, Lawrence Berkeley National Laboratory, Berkeley, CA 94720, USA}
\author{Eli~Rotenberg}
\affiliation{Advanced Light Source, Lawrence Berkeley National Laboratory, Berkeley, CA 94720, USA}
\author{Andrea~Damascelli}
\affiliation{Quantum Matter Institute, University of British Columbia, Vancouver, BC V6T 1Z4, Canada}
\affiliation{Department of Physics \& Astronomy, University of British Columbia, Vancouver, BC V6T 1Z1, Canada}
\author{Johnpierre~Paglione}
\affiliation{Center for Nanophysics and Advanced Materials, Department of Physics, University of Maryland, College Park, Maryland 20742, USA}
\affiliation{Canadian Institute for Advanced Research, Toronto, Ontario M5G 1Z8, Canada}
\date{\today}

\begin{abstract}

The transition metal dichalcogenide TiSe$_2$ has received significant research attention over the past four decades. Different studies have presented ways to suppress the 200~K charge density wave transition, vary low temperature resistivity by several orders of magnitude, and stabilize magnetism or superconductivity. Here we give the results of a new synthesis technique whereby samples were grown in a high pressure environment with up to 180~bar of argon gas. Above 100~K, properties are nearly unchanged from previous reports, but a hysteretic resistance region that begins around 80~K, accompanied by insulating low temperature behavior, is distinct from anything previously observed. An accompanying decrease in carrier concentration is seen in Hall effect measurements, and photoemission data show a removal of an electron pocket from the Fermi surface in an insulating sample. We conclude that high inert gas pressure synthesis accesses an underlying nonmetallic ground state in a material long speculated to be an excitonic insulator.

\end{abstract}

\maketitle

\section{\label{sec:Intro}Introduction}

Titanium diselenide is one of the most studied members of the transition metal dichalcogenide (TMD) family. As with other TMDs, weak van der Waals bonding along the \textit{c}-axis of its hexagonal structure means that relatively minor tweaks to unit cell size, stoichiometry, or interlayer dynamics can have dramatic effects on physical properties. In TiSe$_2$, these changes are most evident in investigation of the charge density wave (CDW) that emerges at 200~K under normal circumstances\cite{DiSalvoTiSe2}. The TiSe$_2$ Fermi surface is not susceptible to nesting, a typical driver of charge ordering in TMDs\cite{RossnagelCDWTMD}, so other explanations have been proposed: different variations of the Jahn-Teller effect\cite{HughesJT, WhangboJT, vanWezelCDWCause} or an excitonic insulator state resulting from a small indirect band gap or overlap\cite{JeromeExcitonicInsulator}. Recent experiment has given backing to the latter scenario\cite{CercellierExcitonicInsulator, KogarMEELS}. Despite this, TiSe$_2$ single crystals show metallic low temperature behavior, with an overall resistivity decrease from room temperature.

Both the application of high pressure and the intercalation or substitution of new atoms to TiSe$_2$ have been used to change the character of the CDW, generally suppressing it and in some cases leading to superconductivity or magnetic ordering\cite{DiSalvoTi1-xVxSe2, KuranovCrFeCoTiSe2,  MorosanCuxTiSe2, KusmartsevaTiSe2Pressure, SasakiFexTiSe2, MorosanPdxTiSe2, ChenTiPtSe2, LuoCrDoping}. In this paper, we present a way of stabilizing new properties in TiSe$_2$ at ambient pressure without the use of additional atoms. By applying up to 180~bar of pressure with argon gas during growth, we have synthesized both single and polycrystals that, below 100~K, exhibit a first-order transition together with a large increase in the resistance and magnitude of the Hall coefficient. We find pressure growth to be fundamentally distinct from substitution, as samples show very similar transport, magnetic, and structural properties to typical TiSe$_2$ at higher temperatures. Instead, the presence of high inert gas pressure reduces selenium vacancy formation, counteracting an extrinsic metallic component in the resistance and allowing for observation of a previously obscured insulating ground state, traces of which have been seen in other studies. Photoemission measurements give evidence for the disappearance of an electron pocket in an insulating sample that is present in semiconducting and metallic ones, in line with this assertion.

\section{\label{sec:Methods}Experimental Methods}

Typically, TiSe$_2$ crystals have been grown by chemical vapor transport (CVT) with excess Se or, more popularly, I$_2$ as the transport agent\cite{DiSalvoTiSe2, KusmartsevaTiSe2Pressure, HildebrandDefectsTiSe2, LioiSeMigration, RiekelNeutron}. In contrast, for this study samples were grown at elevated pressure using argon gas in a Morris HPS-3210 furnace [Fig. 1(a)]. This furnace can reach pressures up to 200 bar at 1000~\degree{}C by introducing Ar into a stainless steel growth chamber. The pressure in the chamber varies in a consistent manner with temperature; the values reported here correspond to the maximum observed pressure for each growth, which (depending on maximum temperature) was about 60-70\% greater than at room temperature.

\begin{figure}
    \centering
    \includegraphics[width=0.49\textwidth]{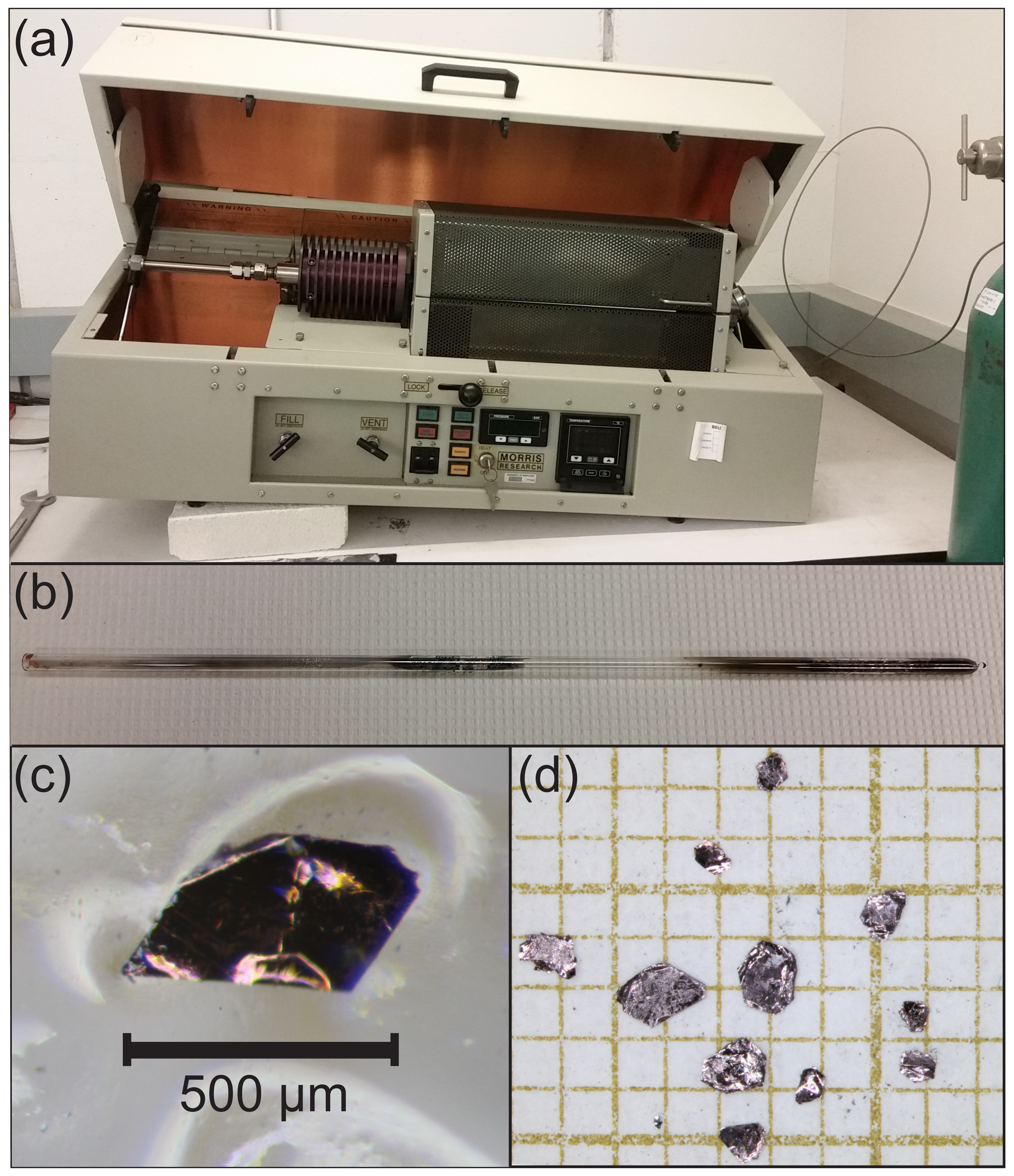}
    \caption{(a) The Morris HPS-3210 furnace used for growth. A quartz ampule, open at one end, was inserted with its closed end on the right hand side near the heating element. The entire chamber was then sealed and the furnace lid closed. Tilting the furnace is suspected to help nucleate crystals. (b) A typical ampule after growth. The dark area on the closed right end is polycrystalline TiSe$_2$, while the region in the middle is elemental Se. (c) A small, fragile, pressure-grown TiSe$_2$ crystal. (d) Larger pressure-grown TiSe$_2$ single crystals (on 1$\times$1~mm$^2$ scale paper).}
    \label{fig:Figure1}
\end{figure}

Traditional CVT is not possible in the pressure furnace, both because of the large amount of Ar gas present and the fact that iodine vapor would damage the chamber, so the actual process was closer to a ``Se flux'' growth. Se shot (99.999+\%, Alfa Aesar) and crushed Ti slugs (99.98\%, Alfa Aesar) were mixed together at the bottom of a quartz ampule with a 40~cm length and 0.75~cm inner diameter. Growths were attempted with Ti:Se ratios ranging from 1:2 to 1:9, and several different temperature sequences. The most frequently used profile was a 1:9 ratio and heating the furnace at 48~\degree{}C/hour to 700~\degree{}C, where it stayed 24-48~hours. The sample space was then slowly cooled at 4.8~\degree{}C/hour to 400~\degree{}C, after which it was passively cooled to room temperature; only then was the chamber returned to ambient pressure. For comparison, we also grew single crystals via CVT with I$_2$ and a flux technique\cite{CanfieldFluxGrowth} using excess Se (in a 9:1 ratio with Ti) in alumina crucibles in about 1/3~atm of Ar gas inside a sealed quartz tube.

All pressure furnace growths produced a large number of polycrystalline chunks of TiSe$_2$, but only about half also resulted in single crystals large enough for transport measurements. There was no identifiable correspondence between growth pressure, temperature profile, or Ti:Se ratio and the successful production of large crystals. Empirically, it seemed that propping up end of the furnace opposite from where the reactants were located at an angle helped to form large single crystals. Doing this may concentrate Se at the end of the ampule where the Ti slugs are located, since the excess Se often condensed further up the length of the tube [Fig.~1(b)]. It could also help amplify any natural temperature gradient in the long, narrow furnace and approximate the conditions for vapor transport.

Resultant single crystals varied in appearance between growths, as seen in Figs.~1(c) and 1(d). Some were small, whispy, and flexible, less than half a millimeter in length and 15-60~$\upmu{}$m in thickness. Others were larger and sturdier, over 1~mm wide and 200~$\upmu{}$m thick. In either case, single crystal x-ray diffraction (XRD) confirmed that the platelike crystals always grew with the $c$-axis out of plane, as would be expected for hexagonal, layered TiSe$_2$. Throughout this paper, single crystals are labeled by the maximum pressure reached during growth, with those from the same batch distinguished by lettering.

Synchrotron powder XRD data were obtained through the 11-BM beamline rapid access mail-in program at the Advanced Photon Source at Argonne National Laboratory, and refinements were made with the GSAS-II software package\cite{TobyGSAS-II}. Single crystal XRD measurements were made on a Bruker APEX2 Diffractometer with Mo K$_{\alpha{}}$ radiation. The integral intensities were corrected for absorption with the SADABS software\cite{KrauseSADABS} using the integration method. The structure was solved with the ShelXS-2015 program and refined with the ShelXL-2015 program and least-square minimization using the ShelX software package\cite{SheldrickSHELXL}. Angle-resolved photoemission spectroscopy (ARPES) measurements were performed at the MAESTRO beamline 7.0.2.1 of the Advanced Light Source in the Lawrence Berkeley National Laboratory, where samples were cleaved and measured at 20~K with a base pressure of $2\times 10^{-11}$~Torr. Photoelectrons were detected by a Scienta R4000 analyzer equipped with a deflector, where the energy and angular resolution were better than 20~meV and 0.2\degree{}, respectively. Electrical transport measurements were carried out in 9~T and 14~T Quantum Design Physical Properties Measurement Systems, and a 14~T Quantum Design DynaCool. The 14~T PPMS and DynaCool were also used for heat capacity measurements. Magnetization was measured using the DynaCool vibrating sample magnetometer as well as two versions of the 7~T Quantum Design Magnetic Properties Measurement System, the MPMS XL and MPMS3.

\section{\label{sec:Physical Properties}Physical Properties}

\subsection{\label{sec:XRD}Structural Characterization}

\begin{figure}
    \centering
    \includegraphics[width=0.49\textwidth]{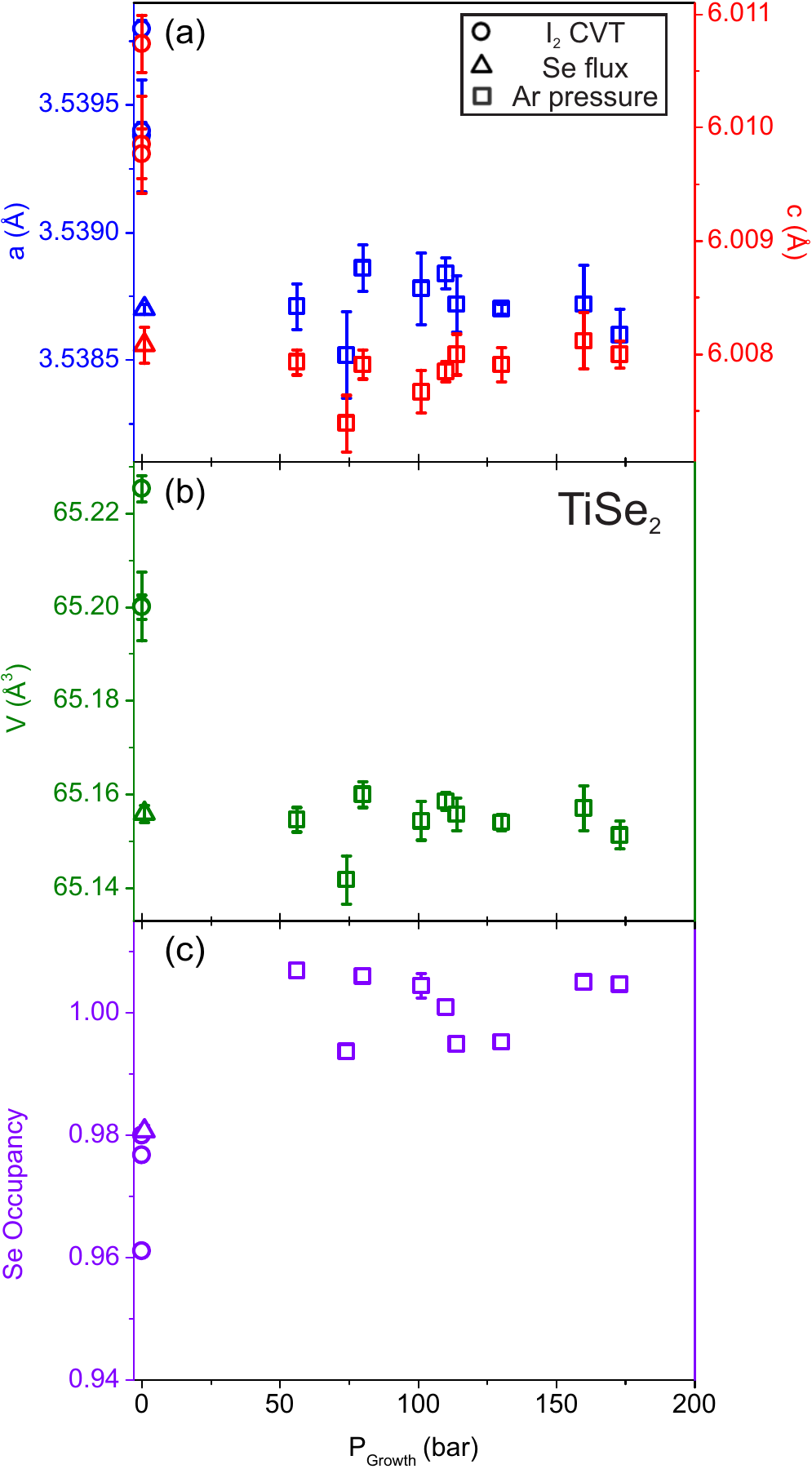}
    \caption{TiSe$_2$ structural data obtained from refinement of powder synchrotron data of CVT (squares, ``0~bar''), Se flux (triangles, ``1~bar''), and pressure (circles) growths. (a) shows the \textit{a}- (left, blue) and \textit{c}- (right, red) axis lengths and (b) the volumes, with error bars magnified ten times from those given by GSAS-II. Note that data points for two of the CVT growths nearly completely overlap. (c) is the refined Se occupancy of the same data, with error bars, where a CVT and Se flux growth overlap at 0.98.}
    \label{fig:Figure2}
\end{figure}

Synchrotron diffraction patterns were taken at 295~K of ground polycrystals from 13 growths: three separate CVT growths with temperatures of either 550 or 575 $\degree{}$C at the hot end of the ampule (thought to be optimal for CVT)\cite{DiSalvoTiSe2}, a Se flux growth, and nine pressure growths with maximum pressures in the range 56-173~bar. Figure~2 shows the (a) lattice parameters, (b) volumes, and (c) Se site occupancies for all data sets, where vapor transport is denoted as ``0~bar'' of growth pressure, Se flux as ``1~bar'', and growth methods are further distinguished by symbols. Values for all samples are very close to those reported previously\cite{RiekelNeutron, MorosanCuxTiSe2}. Elemental Se was present in some cases, unsurprising given the use of excess in the growth. There is a very small but nevertheless consistent difference in unit cell size and composition between the synthesis techniques. Both Se flux and pressure growths have smaller lattice parameters, perhaps attributable to the absence of larger iodine atoms that might replace Se, or the result of slight lattice compression due to the higher pressure during nucleation. This shift is much smaller than that brought about by small amounts of chemical dopants or intercalants\cite{KuranovCrFeCoTiSe2, LuoCrDoping, MorosanCuxTiSe2}. Among pressure-grown samples, there is again no trend in lattice size with the amount of Ar pressure applied.

There is also a clear difference in Se occupancy between pressure growth and the other two methods. The presence of Se vacancies is known to be highly sensitive to growth method and temperature, and can impact transport behavior\cite{DiSalvoTiSe2, HuangTiSe2Vacancies}. Vapor transport and Se flux result in vacancies of 2-4\%, while those for pressure growth are all within 1\% of full occupation. An explanation for this is that the presence of a significant amount of inert gas suppresses the high vapor pressure of Se and as a result reduces vacancy formation. In contrast, vapor transport relies on reaction in the gas phase with a third element, iodine, making it more susceptible to reduced Se content or site substitution. Se flux avoids this issue, but with a low argon pressure that does not sufficiently combat the volatility of Se. Comparable temperatures and amounts of excess Se were used in flux and pressure growths, thus it is evident that higher pressure is the key factor to reduced vacancies. Significant differences in the properties of pressure-grown samples compared to Se flux or I$_2$ CVT that we will show further on are then attributable to Se occupancy, rather than a slightly smaller lattice.

\begin{table}
	\centering
    \caption{Parameters obtained from single crystal XRD at 150~K. Here and in subsequent figures, single crystals from the same batch (and therefore with the same P$_{\text{Growth}}$) are distinguished by lettering.\\}
    \label{tab:Table1}

\begin{tabular}{ | c | c | c | c | c | c | c |}
	\hline
	P$_{\text{Growth}}$ (bar) & \textit{a} (\AA{}) & \textit{c} (\AA{}) & $wR_2$ \\
	
	\hline
	101A &    ~3.5415(11)~    &    ~6.0198(19)~    &   ~0.0494~   \\
	\hline

	101B &    ~3.5355(9)~    &    ~6.0112(16)~    &   ~0.0556~   \\
	\hline
	
	114A &    ~3.5332(15)~    &    ~6.007(3)~    &   ~0.0553~   \\
	\hline
	
	138 &    ~3.5432(10)~    &    ~6.0195(17)~    &   ~0.0615~   \\
	\hline

	140 &    ~3.5275(11)~    &    ~5.9969(19)~    &   ~0.0509~   \\
	\hline
	
\end{tabular}

\end{table}

Five single crystal samples grown in the range 101-140~bar, including two from the same batch, were also selected for XRD at 150~K [Table~I]. These data are not directly comparable to room temperature values as they are determined by a combination of thermal contraction and CDW-related lattice distortion. Nevertheless, they demonstrate that T$_{\text{CDW}}$ is above 150~K for these samples, in spite of different growth conditions and low temperature resistivities 2-5 times higher than the 300~K value. Using the reported TiSe$_2$ lattice parameters\cite{RiekelNeutron} and thermal expansion coefficients\cite{WiegersThermalExpansion} we can estimate 150~K lattice parameter values, before accounting for the effect of the CDW on the lattice, to be \textit{a}~=~3.530~\AA{} and \textit{c}~=~5.993~\AA{} for typical TiSe$_2$. With one exception, we see 150~K lattice parameters that are similar to or larger than room temperature values. The reason for this is charge ordering, which by 150~K can result in a distortion of more than 10$^{-2}$~\AA{} in typical TiSe$_2$\cite{DiSalvoTiSe2, MonneyLattice}. While applied pressure is known to suppress the CDW\cite{KusmartsevaTiSe2Pressure}, pressure growth evidently does not, since at 150~K samples have experienced a lattice expansion.

\subsection{\label{sec:Transport}Electrical Transport}

The temperature-dependent resistance of the single crystals from Table I is shown in Fig.~3, with resistances scaled to 300~K values. The behavior shown is representative of what is seen in a larger number of samples that we have measured. Above 100~K, the behavior does not differ from previous reports on TiSe$_2$. Following convention we used the kink in the derivative (a peak in the second derivative) to identify the onset of charge ordering\cite{DiSalvoTiSe2}. The values we see for all samples are just above 200~K, the same as standard TiSe$_2$, and concordant with the conclusion from 150~K single crystal XRD. All samples have a rise upon cooling at 200~K regardless of behavior at lower temperatures. On the other hand, the peak in resistance, normally centered around 165~K, comes at a lower temperature in most of our samples. A similar effect has been reported in samples grown without iodine\cite{TaguchiI2Free}, or at higher temperature\cite{DiSalvoTiSe2, HuangTiSe2Vacancies}. The ratio R$_{\text{peak}}$/R(300~K) for our samples can be as high as six, larger than has been achieved with CVT crystal growth\cite{HuangTiSe2Vacancies, TaguchiI2Free}. An increased peak height (relative to 300~K resistivity) has previously been interpreted as signifying fewer Se vacancies and correspondingly higher crystal quality\cite{DiSalvoTiSe2, CravenTiSe2HC, HildebrandDefectsTiSe2}. Another recent paper argued that Se deficiency is actually beneficial to charge ordering, a conclusion based purely on the size of the resistance increase in the 150-200~K region\cite{HuangTiSe2Vacancies}. In comparing multiple growth methods, pressure-grown crystals show a taller peak, but at a lower temperature, and without Se deficiency. From this we believe that there is no correspondence between peak height, the temperature at which it occurs, and crystal quality. The broadness of the peak, and the fact that it comes 30-50~K lower than T$_{\text{CDW}}$, means that it likely represents simply a change in dominant scattering mechanism within the charge ordered phase, and that its specifics are not as significant. This is supported by the fact that our own samples with similar peak heights show differing behavior upon further cooling.

\begin{figure}
    \centering
    \includegraphics[width=0.45\textwidth]{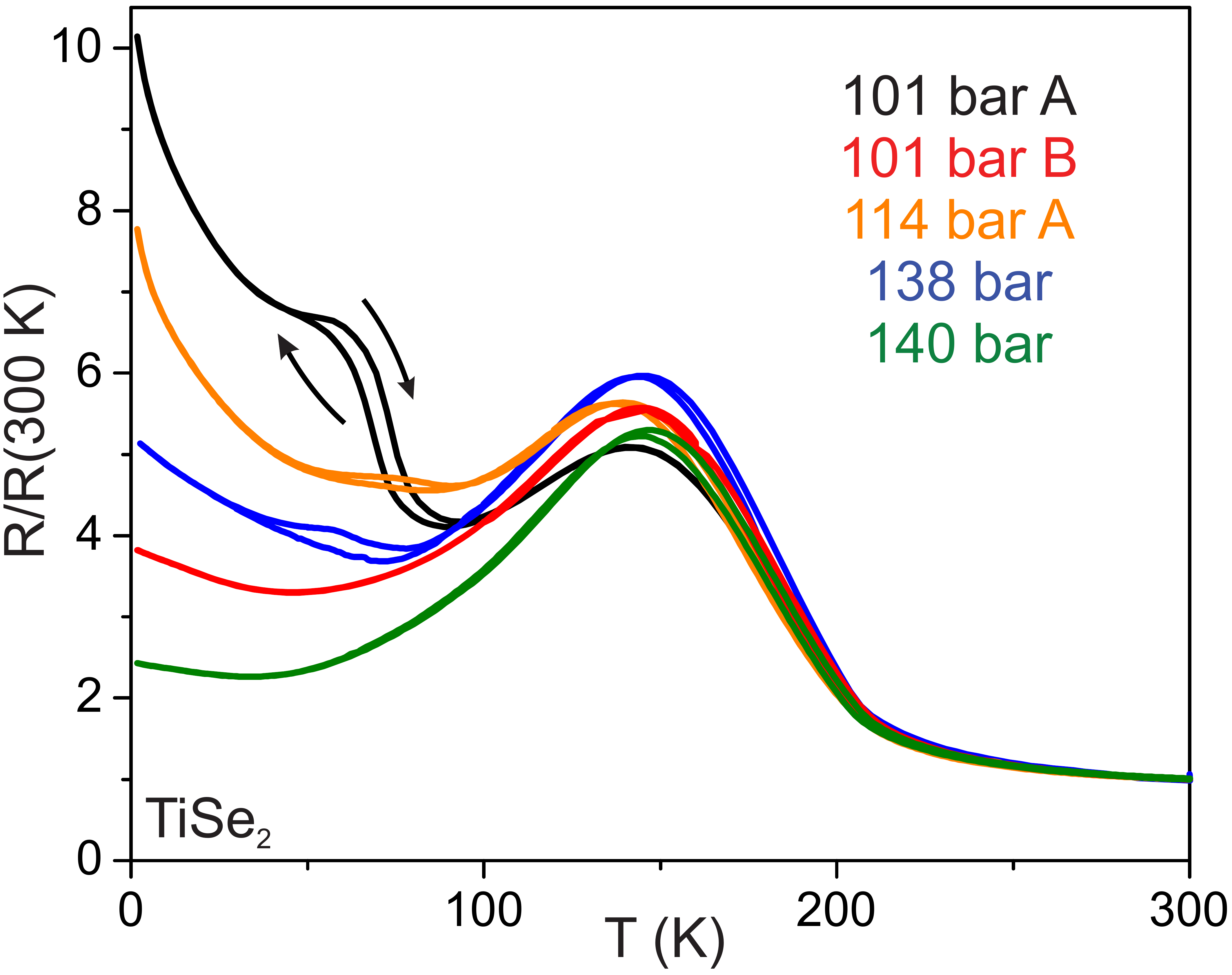}
    \caption{Resistance (scaled to 300~K value) as a function of temperature for pressure-grown TiSe$_2$ single crystals listed in Table I, with maximum growth pressures noted. Note that even the two samples from the same batch exhibit very different behavior. Hysteresis in the 30-80~K region is marked by arrows for the black curve to show the difference in warming and cooling, which is the same for unmarked samples. In some cases there is also hysteresis around the CDW-related upturn from 150-200~K.}
    \label{fig:Figure3}
\end{figure}

The more significant departure from previously observations comes at lower temperatures, where many pressure-grown samples show a large increase in resistivity. Additionally, temperature hysteresis often opens around 80~K in a similar range to where insulating character emerges, before closing near 30~K. Visible in Fig.~3, this is emphasized in Fig.~4, which shows $\Delta{}R{}$, defined as the difference in resistance value between warming and cooling, scaled to its room temperature value, for five samples from a single 101~bar growth. Except for a single nonhysteretic sample (the green curve), the greatest difference occurs between 30-100~K, with samples also showing appreciable $\Delta{}R{}$ near and above the higher temperature resistance peak. This effect is still present when temperature is swept slowly or stabilized at each data point, and so is not a result of temperature lag or sample heating. In some cases there is a noticeable kink at the hysteresis opening, further confirmation that the transition is more than just measurement error. Application of fields up to 140~kOe does not affect overall behavior. Polycrystalline samples were always insulating and hysteretic, often to a far greater degree, but we present only single crystal resistance data throughout this paper in order to demonstrate that the effect is inherent to pressure-grown TiSe$_2$ and not a result of insulating impurities, grain boundaries, or other effects that make polycrystalline transport measurements less reliable.

Not all single crystals show the 80~K transition; some behave like typical, metallic, CVT-grown TiSe$_2$, and behavior below 100~K varied even for samples from the same growth [Fig.~4, inset; note the logarithmic y-axis]. The height of the CDW peak is also inconsistent and its temperature can vary by as much as 10~K. The lowest growth pressure to produce single crystals was 56~bar, and the highest was 140~bar, but polycrystals were grown at maximum pressures of 10-180~bar and were more universally insulating. As with lattice parameters, there is no clear link between a specific growth pressure and resistivity behavior (or any other measurable quantity). Instead, it seems that pressure synthesis can stimulate this behavior, but does not guarantee it. The difference between 50 and 200~bar of pressure is likely insignificant, since in any case the pressure is much higher than during CVT or flux growth.

A comparison of $\rho{}$(T) for crystals grown by I$_2$ CVT, Se flux, and Ar pressure [Fig.~5(a)] makes clearer the differences in resultant single crystals. All show the CDW-associated rise in $\rho$ at the same temperature. The CVT crystal has a peak at 165~K and $\rho$(1.8~K)~$<$~$\rho$(300~K). The Se flux sample has a larger peak, suppressed in temperature to 150~K, and slightly higher resistivity at 1.8~K than room temperature. A pressure-grown crystal has a local maximum at even lower temperature (140~K) with a comparable size (relative to $\rho{}$(300~K)) and a hysteretic, insulating transition. T$_{\text{CDW}}$ has typically been identified as the beginning of the flat minimum region in the first derivative [Fig.~5(b)], which corresponds to the onset of charge ordering in neutron measurements\cite{DiSalvoTiSe2}. However, the minima for Se flux and pressure-grown samples are influenced by the suppression of T$_{\text{peak}}$. We therefore used the peak in the second derivative, equivalent to the kink in the first derivative, to define T$_{\text{CDW}}$. In the inset to Fig.~5(b) it is clear that this occurs at the same temperature for all three samples, and in fact the CVT and pressure-grown samples have nearly identical first and second derivatives above 200~K. While both Se flux and pressure-grown crystals have an elevated low temperature resistance compared to CVT samples, pressure growth is further distinguished by the more significant insulating behavior and temperature hysteresis.

As with longitudinal resistivity, the Hall resistance [Fig.~6(a)] in pressure-grown crystals above 200~K is generally similar to typical TiSe$_2$\cite{DiSalvoTiSe2}. The Hall coefficient R$_{\text{H}}$ is initially positive and crosses zero at 150-170~K [Fig.~6(a), inset], which like the $\rho$(T) peak is slightly lower than our own CVT crystals and previous reports\cite{DiSalvoTiSe2, TaguchiI2Free}. Elemental substitution leads to a more substantial temperature suppression\cite{LevyTiVSe2, LevyImpurities}. Some pressure-grown showed a slight increase in R$_{\text{H}}$ just before the zero-crossing not seen with vapor transport. At low temperatures R$_{\text{H}}$ can reach large negative values up to two orders of magnitude larger than those of typical TiSe$_2$\cite{DiSalvoTiSe2} and an order of magnitude above those measured even for insulating Ti$_{1-x}$M$_x$Se$_2$ (M~=~As, Sc, Nb, Ni, Re, or Y)\cite{LevyImpurities}, indicating that a reduced carrier concentration, rather than impurity scattering, is the reason for increased resistance. Generally, samples with more insulating low temperature behavior had a larger $\lvert{}$R$_{\text{H}}\rvert{}$. The Hall signal becomes more linear with decreasing temperature below the R$_{\text{H}}$ sign change, indicating transport dominated by a single electron band despite the presence of multiple carrier types at higher temperatures. Overall the Hall data support the idea of a change in band structure at low temperatures between pressure and CVT-grown TiSe$_2$. Another thing we noticed in measurements was an 80~K maximum in $\lvert{}$R$_{\text{H}}\rvert{}$ for CVT crystals. Extrema in R$_{\text{H}}$ at a similar temperature have been seen in other intrinsic\cite{DiSalvoTiSe2} and metal-doped\cite{LevyImpurities} samples. Given that this is the same temperature as the beginning of hysteresis, it suggests that pressure growth emphasizes or strengthens some phenomenon already present in other forms of TiSe$_2$.

\begin{figure}
    \centering
    \includegraphics[width=0.49\textwidth]{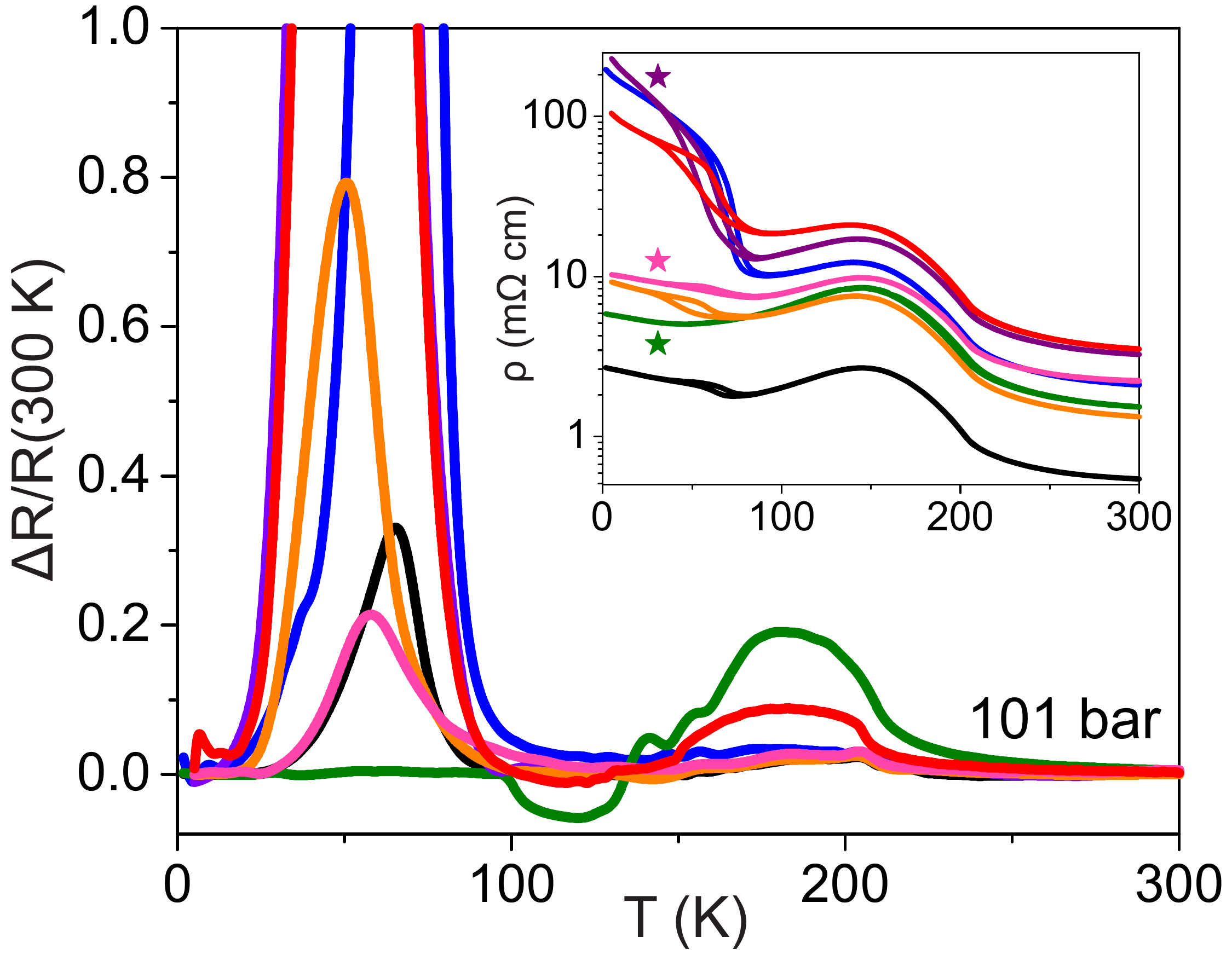}
    \caption{$\Delta{}R \equiv{} R_{warming} - R_{cooling}$, scaled to room temperature resistance, as a function of temperature for seven crystals from the same 101~bar growth. Hysteresis is most evident around 80~K but also manifests at the higher temperature CDW resistance peak. $\Delta{}R/R$ reaches a maximum of 5.0, 6.7, and 6.9 for the red, purple, and blue curves, respectively. Inset: $\rho{}$(T) of each sample, with matching colors. Stars indicate samples for which data are also presented in other figures: 101B (Table I and Fig.~3, here green), 101C (Fig.~5, pink), and 101D (Fig.~7, purple).}
    \label{fig:Figure4}
\end{figure}

\begin{figure}
    \centering
    \includegraphics[width=0.49\textwidth]{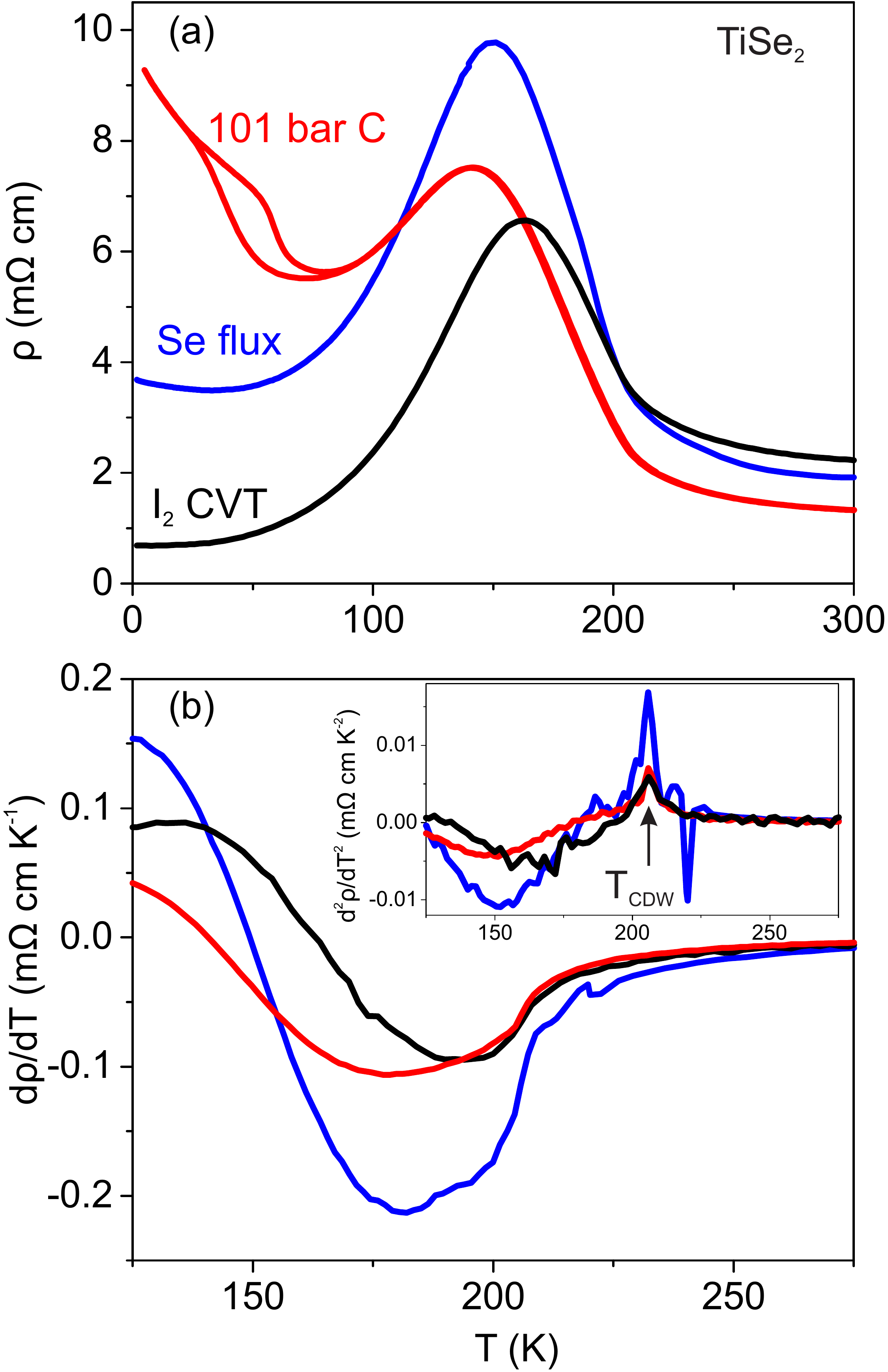}
    \caption{(a) Temperature-dependent resistivity for TiSe$_2$ single crystals grown by I$_2$ chemical vapor transport, excess Se flux, and 101~bar of Ar gas pressure. At high temperatures the behavior of all three samples is similar, but differences emerge below 150~K. No hysteresis is observed in the CVT or flux crystals. (b) The derivative of the cooling data from (a) in a narrower temperature region. The inset is the second derivative of the same data, where the peak is identified as the onset of the CDW.}
    \label{fig:Figure5}
\end{figure}

\subsection{\label{sec:HC}Heat Capacity}

Heat capacity measurements were taken from 300~K to 2~K on a polycrystalline chunk of TiSe$_{2}$ grown at 160~bar [Fig.~6(b)], which had shown an insulating transition in transport measurements. We observe no features in the corresponding temperature range, and the shape and values of the data are very similar to what has been measured before\cite{CravenTiSe2HC}. The lack of a feature in the hysteretic region is not wholly surprising, as even that corresponding to the higher temperature CDW is subtle. Low temperature measurements on the same polycrystal and two others grown at different pressures are shown in the inset to Fig.~6(b). All data fit well to the standard specific heat equation $C/T = \gamma{} + \beta{}T^{2}$. In this equation, $\gamma$ is the Sommerfeld coefficient and $\beta$ can be used to calculate the Debye temperature $\theta{}_{D} = (\frac{12\pi{}^{4}N_{A}k_{B}n}{5\beta{}})^{\frac{1}{3}}$, where $N_{A}$ is the Avogadro number, $k_{B}$ Boltzmann's constant, and $n = 3$ the number of atoms per formula unit. Results were similar for all three samples. The computed $\gamma$ values are small: 0.14, 0.16, and 0.19~$\frac{\text{mJ}}{\text{mol~K}^2}$ respectively for the 98, 130, and 160~bar samples, reflecting the small low temperature density of states at the Fermi energy. $\theta{}_D$ is 220, 244, and 209~K for the same data. The reference values for TiSe$_2$ powder\cite{CravenTiSe2HC} are $\gamma$~=~0.19~$\frac{\text{mJ}}{\text{mol~K}^2}$ and $\theta_{D}$~=~251~K, both similar to but slightly higher than those derived from pressure-grown samples.

\begin{figure*}
    \centering
    \includegraphics[width=1\textwidth]{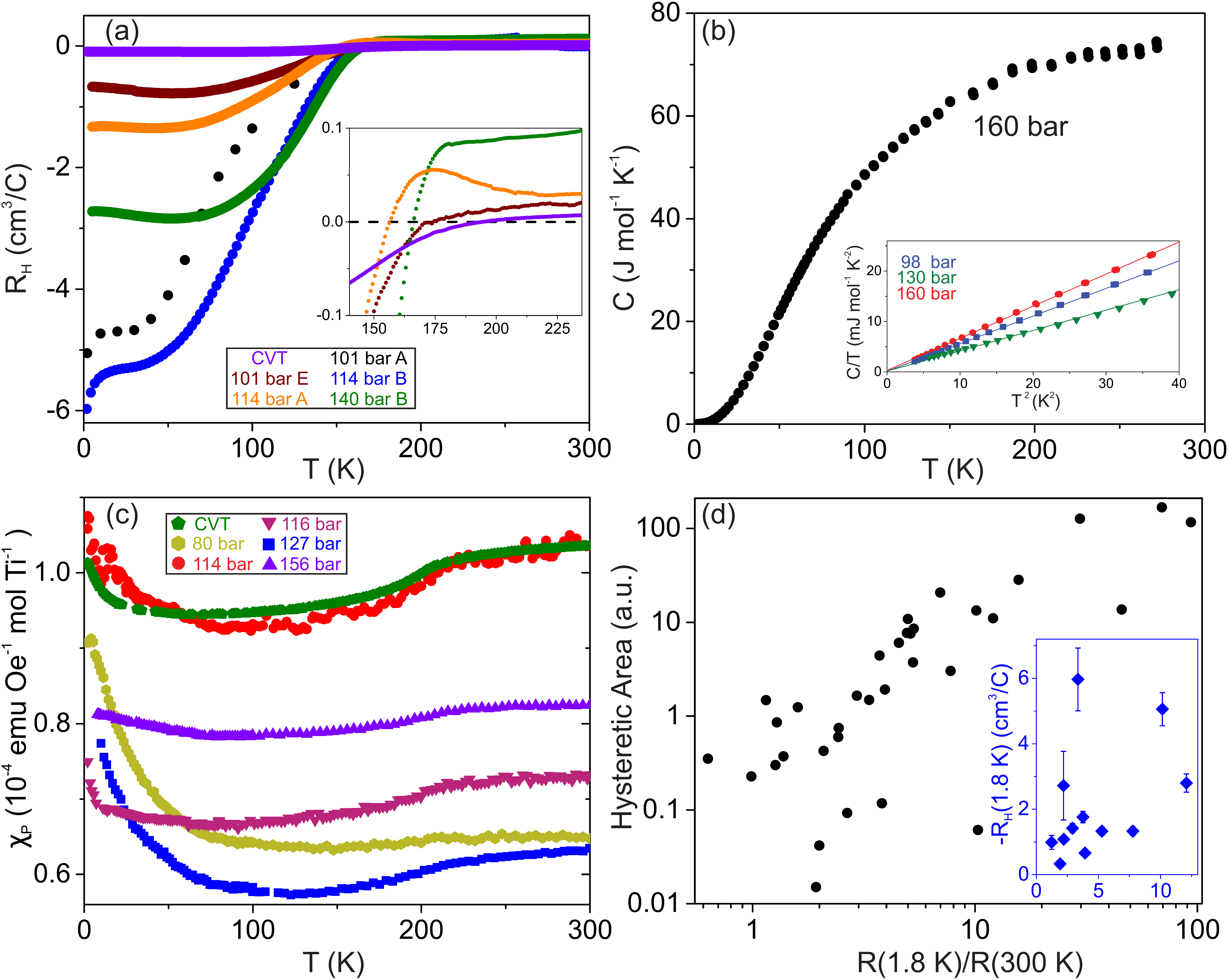}
    \caption{(a) The temperature dependence of the Hall coefficient of TiSe$_2$ single crystals. Note that two CVT samples, with nearly identical data and changes too small to see on this scale, are included. The inset shows the zero-crossing region for selected samples from the main plot. (b) Specific heat of a polycrystalline TiSe$_2$ sample with a maximum growth pressure of 160 bar as a function of temperature. Inset: Low temperature data for the same sample and two other polycrystals, with maximum growth pressures noted. Lines are fits to the Debye low temperature specific heat model. (c) Paramagnetic susceptibility of polycrystalline TiSe$_2$ chunks taken in constant applied fields of 20 (116~bar), 50 (80, 127, and 156~bar), or 70~kOe (CVT and 114~bar) due to the small intrinsic moment. (d) The hysteretic area (see Discussion for details) versus resistance increase for pressure-grown single crystals, on a log-log scale. Inset: the base temperature Hall coefficient as a function of the same quantity. More insulating samples generally showed a larger $\lvert{}$R$_{\text{H}}\rvert{}$. Error bars come from uncertainty in the measurement of sample thickness and in some cases are smaller than the symbol size.}
    \label{fig:Figure6}
\end{figure*}

\subsection{\label{sec:Magnetization}Magnetic Susceptibility}

Like heat capacity, magnetization measurements for multiple growths differed little from vapor transport-grown TiSe$_2$. The total magnetic susceptibility is small, on the order of 10$^{-6}$~emu~(Oe~mol~Ti)$^{-1}$ , since the paramagnetic and diamagnetic components are comparable in magnitude\cite{DiSalvoTiSe2, MorosanCuxTiSe2, MorosanPdxTiSe2}. Fig. 6(c) shows the paramagnetic susceptibility $\chi{}_P$ after subtraction of the core diamagnetic contribution\cite{BainDiamagCorrxn}. The shape matches results from our own CVT-grown samples and what has been presented in the past\cite{DiSalvoTiSe2}, and in fact has a similar shape to d$\rho$/dT. We attribute the rise at lowest temperatures in some curves to paramagnetic impurities. Data were taken under large fields ($\geq 20 $~kOe) to enhance the weak signal, but curves had the same appearance over the range 0-140~kOe.

\begin{figure*}
    \centering
    \includegraphics[width=.95\textwidth]{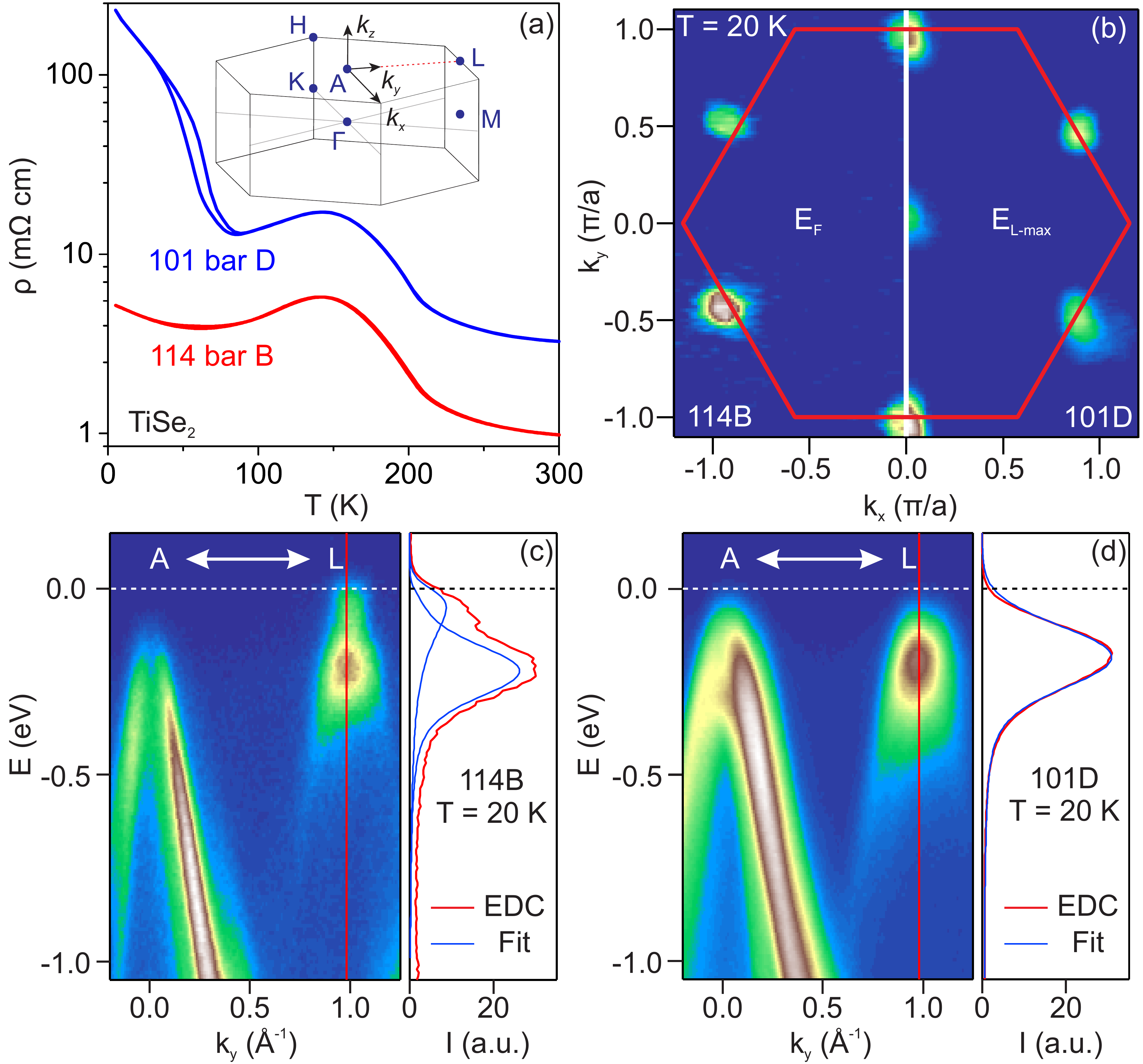}
    \caption{(a) Resistivity as a function of temperature for the two pressure-grown ARPES samples, with a schematic drawing of the Brillouin zone. 114B shows a small amount of hysteresis that is less obvious with the logarithmic scale. (b) Comparison of the constant energy contour measured at $k_z=\pi/c$ (the A\textendash{}L\textendash{}H plane) at E$_{\text{F}}$ for 114B (left) and the L valence band maximum (60~meV below E$_{\text{F}}$) for 101D (right) at T~=~20~K, plotted in the $k_x < 0$ and $k_x > 0$ half-plane, respectively. (c) and (d) ARPES map measured along the A\textendash{}L cut (the dashed red line in (a)) for 114B and 101D, respectively. To the right of each cut are the EDCs at the L-point (vertical red solid line), which were fit by Voigt functions multiplied by the Fermi-Dirac distribution(blue curves).}
    \label{fig:Figure7}
\end{figure*}

\subsection{\label{sec:ARPES}ARPES}

To understand the origin of the insulating ground state in some crystals, we performed comparative ARPES measurements on two crystals, one semimetallic (114B in Fig.~7(a)) and another more insulating (101D). Measurements were made at 20~K, below the onset of hysteretic resistance. Previous studies have shown that the Fermi surface in the CDW phase is composed of a hole pocket at the $\Gamma$-point (Ti-$3d$) and an electron pocket at the L-point (Se-$4p$)\cite{VydrovaSoftARPES, ChenSingleLayer}. Despite discussions of the semimetallic versus semiconducting nature of TiSe$_2$ above the CDW transition\cite{RaschSemimetalSemiconductor}, the previous consensus was that an electron pocket crosses the Fermi level at the L-point in the CDW phase. However, we find that the more insulating sample presents neither an electron nor hole pocket anywhere in k-space at the Fermi energy. Fig.~7(b) compares the constant binding energy contours measured in the A-L-H plane ($k_z=\pi/c$), where data for 114B (at E$_{\text{F}}$) and 101D (60~meV below E$_{\text{F}}$), at the maximum of the L-point valence bands) are shown in the left and right half planes, respectively. The Fermi surface of 114B is composed of electron pockets at the L-point, whereas in the more insulating 101D the states closest to the Fermi level are the hole bands at the $\Gamma$- and L-points at a larger binding energy. This is most evident from the measured band dispersion along the A\textendash{}L direction (the red dotted line in the Fig.~7(a) inset). For 114B [Fig.~7(c)], a small electron pocket crosses the Fermi level at $k_y~\approx 1.0~$\AA{}$^{-1}$. The energy distribution curve (EDC) at this momentum (marked by solid red line) is shown in the right panel of Fig.~7(c), where it is fit with two bands, consistent with previous reports\cite{ChenSingleLayer}. This is distinct from 101D, shown in Fig.~7(d), where no band crosses the Fermi level and the EDC can be fit with just a single Voigt distribution multiplied by the Fermi-Dirac function. The lack of any quasiparticle band crossing the Fermi energy would naturally explain the insulating behavior of 101D and should also result in a reduced carrier concentration. Scans throughout the entire $k_z$ dispersion confirmed the absence of any intensity at E$_{\text{F}}$ at the Brillouin zone edge.

\section{\label{sec:Discussion}Discussion}

Above 100~K, there is little to distinguish TiSe$_2$ grown with iodine vapor transport or at high argon pressure. The difference between the two of the sudden decrease in $\chi{}$(T) and the maximum in d$^{2}\rho{}$/dT$^{2}$ (the d$\rho{}$/dT inflection point), both associated with the onset of charge ordering,\cite{DiSalvoTiSe2, LevyTiVSe2} is 5~K or less. XRD data show that the lattice of pressure-grown samples is about 0.1\% smaller than for CVT. However, pressure growth results in no appreciable selenium vacancies, compared to about 2-4\% reduced stoichiometry in CVT or flux samples. While small, these changes are very consistent. The most noteworthy aspect of pressure growth is that it can lead to insulating and hysteretic low temperature behavior. And although TiSe$_2$ becomes more insulating with $\leq{}$5\% V doping\cite{DiSalvoTi1-xVxSe2, LevyTiVSe2} or the intercalation of Cr, Fe, and Co\cite{KuranovCrFeCoTiSe2, ChenTiPtSe2}, those samples do not display temperature hysteresis. Furthermore, they show CDW temperature suppression\cite{SasakiFexTiSe2}, antiferromagnetic or Curie-Weiss behavior\cite{KuranovCrFeCoTiSe2}, and a more significant change in room temperature lattice parameters\cite{KuranovCrFeCoTiSe2, LuoCrDoping}, distinctly different effects than pressure growth.

The introduction of Pd and Pt, in contrast, do have similarities to our findings. In Pd$_x$TiSe$_2$ with \textit{x}~$\leq{}$~0.03, a second inflection point in d$\rho$/dT occurs near 80~K and transmission electron microscopy shows a strengthened CDW at lower temperature\cite{MorosanPdxTiSe2}. However, the \textit{a}- and \textit{c}-axes are both larger than in unintercalated samples, and further Pd intercalation leads to metallic behavior and superconductivity. Up to \textit{x}~=~0.13, resistivity increases by eight orders of magnitude at low temperatures in Ti$_{1-x}$Pt$_x$Se$_2$, while 300~K lattice constants and T$_{\text{CDW}}$ hardly change\cite{ChenTiPtSe2}. Transport measurements and density functional theory calculations on Pt-doped samples attribute insulating behavior to an increased energy gap. Similarly, our ARPES results demonstrate that pressure growth can lead to the disappearance of an electron pocket. However, hysteresis is not noted with either Pd or Pt. 

TiSe$_2$ has shown inconsistent transport behavior, with I$_2$ CVT-grown single crystals having an overall decrease in resistivity from room temperature\cite{DiSalvoTiSe2, HuangTiSe2Vacancies} and polycrystals being more semimetallic\cite{MorosanCuxTiSe2, ChenTiPtSe2, MorosanPdxTiSe2}. Our own CVT or Se flux crystals can similarly differ in low temperature properties [Fig.~5(a)]. Pressure-grown samples have a decrease in low temperature carrier concentration that is more dramatic in more insulating samples [Fig.~6(d),~inset]. This also seems to be connected to the hysteretic region. In Fig.~6(d) we plot the ``hysteretic area'' of single crystal samples against their scaled resistance increase, R(1.8~K)/R(300~K). This quantity is defined as the area under the $\Delta$R/R(300~K)~vs.~T curve (like those shown in Fig.~4) between 30 and 80~K. The correlation spans several orders of magnitude between more insulating behavior and more pronounced hysteresis, even after scaling the raw resistance.

ARPES results give more insight, showing that a very insulating crystal does not have the electron pocket at the Fermi surface seen in a semiconducting one and previous reports on CVT samples. As ARPES is unable to probe above E$_{\text{F}}$, the exact change to the conduction band cannot be determined: there may be a downward shift of the chemical potential, an increase in the conduction-valence band gap at L, orbital-dependent distortion, or a combination of all three. However, we note that Se vacancies would contribute electron carriers to the system. Reducing those vacancies via pressure growth would be equivalent to hole doping. We speculate that reduced vacancy formation lowers the chemical potential, has also been shown to occur with growth of Bi$_2$Se$_3$ in the same furnace\cite{SyersBi2Se3}. It may also change the size of the gap, similar to the suspected effect of Pt doping. This naturally results in a decreased carrier concentration and increased resistivity at low temperatures.

The fragility of TiSe$_2$'s structure in both real and momentum space means that subtle changes, such as a slight change in E$_{\text{F}}$ or the gap size, will be magnified when transposed onto other properties. The small gap and the variability of a Fermi energy change would also explain the spread of semiconducting-insulating behavior in different samples, emphasized in Fig.~6(d), where even samples from the same batch show different transport behavior. The two ARPES samples have different Fermi surfaces, after all, but were grown under similar pressures. Variation could come from temperature gradients in the furnace, which could approximate the conditions for vapor transport, or other factors more difficult to observe and control during the growth process. Given the consistency of diffraction and transport measurements on polycrystals, it seems likely that the majority of pressure-grown material has reduced vacancies, with fluctuations among individual crystals. Pressure growth of Bi$_2$Se$_3$ had similar variability in resistive behavior and carrier concentration\cite{SyersBi2Se3}.

Anomalies at 80~K have appeared in previous studies of TiSe$_2$. The 200~K commensurate CDW (CCDW) can be suppressed with Cu intercalation or applied pressure, the CDWs in these samples have been shown by x-ray scattering and scanning tunneling microscopy to be incommensurate (ICDWs) in some regions of the phase diagram above the induced superconducting transition\cite{YanTiSe2ICDW, KogarTiSe2ICDW, JoeTiSe2ICDW}. The change in ordering vector is first seen in the 65-80~K range, very close to where hysteresis first emerges in pressure-grown samples, and under pressure the ICDW-CCDW transition is weakly first order. The usual transition of TiSe$_2$ directly to a CCDW is actually atypical for a TMD. TaS$_2$, for example, has three progressively more insulating CDW transitions, with the two at lower temperature being hysteretic\cite{ThompsonTaS2, SiposTaS2}. They correspond to the onset of (with decreasing temperature) incommensurate, nearly commensurate, and fully commensurate charge order. The consistent onset temperature of the ICDW in TiSe$_2$ in comparison to the continuous suppression of the CCDW with pressure or Cu intercalation has led to speculation that there is an inherent mechanism for lower temperature charge ordering that is ``boosted'' to 200~K by excitonic interactions\cite{KogarTiSe2ICDW}. A transition in a similar temperature region was suggested for Pd$_x$TiSe$_2$ with \textit{x}~$\leq$~0.03, where samples are insulating and T$_{\text{CDW}}$ unchanged\cite{MorosanPdxTiSe2}. Even the $\lvert{}$R$_{\text{H}}\rvert{}$ maximum at 80~K for CVT samples supports the notion of an underlying feature at that temperature. The two transitions we see in pressure-grown TiSe$_2$ crystals may signify that pressure growth allows for the observation of both the ``natural'' and ``boosted'' CDWs, perhaps with differing wavevectors, where the former is otherwise obscured by the effects of nonstoichiometry.

\section{\label{sec:Conclusion}Conclusion}

We have shown that TiSe$_2$ crystals grown under argon gas pressures of 10-180~bar can have a much larger resistance and reduced carrier concentration at low temperatures. Synchrotron data show a smaller lattice and reduced Se vacancy formation compared to vapor transport growth, and there is evidence for the elimination of an electron pocket at the Fermi level from photoemission. Prior examples in which the introduction of new atoms caused a low temperature resistance increase lack the hysteretic behavior that we have observed starting around 80~K. We suspect this new behavior stems from an enhancement of charge ordering and suppression of the metallic behavior, attributable to selenium vacancies, that dominates transport behavior in crystals grown by vapor transport. This first order transition may be a signature of a true charge-ordered, excitonic insulator ground state in TiSe$_2$, that has been hinted at in work with applied pressure\cite{JoeTiSe2ICDW}, Cu intercalation\cite{KogarTiSe2ICDW, YanTiSe2ICDW}, or Pd doping\cite{MorosanPdxTiSe2}.

The association between changes to charge ordering in TiSe$_2$ and superconductivity, demonstrated by the observation of CDW incommensuration near a quantum critical point in pressurized or Cu-intercalated samples, is a reason to further explore the possibilities of high pressure crystal growth. Due to their weak interlayer bonding, applied pressure or chemical substitution can significantly impact the behavior of transition metal dichalcogenides. Pressure growth also presents a method of manipulating band gaps in bulk materials without the introduction of extrinsic atoms. This and other aspects of high pressure synthesis can alter observed properties and lead to new discoveries related to TMDs or the many other materials with unstable lattice configurations.

\section{\label{sec:Acknowledgments}Acknowledgments}

This work was supported by Air Force Office of Scientific Research award no.~FA9550-14-1-0332, National Science Foundation Division of Materials Research award no.~DMR-1610349, and the Gordon and Betty Moore Foundation's EPiQS Initiative through grant no.~GBMF4419. D.J.C. was funded in part by the U.S. Department of Energy (DOE), Office of Science, Office of Workforce Development for Teachers and Scientists, Office of Science Graduate Student Research program, administered by the Oak Ridge Institute for Science and Education for the DOE under contract no.~DE‐SC0014664. The work at UBC was undertaken thanks in part to funding from the Max Planck-UBC-UTokyo Centre for Quantum Materials and the Canada First Research Excellence Fund, Quantum Materials and Future Technologies Program. It was also supported by the Killam, Alfred P. Sloan, and Natural Sciences and Engineering Research Council of Canada's (NSERC's) Steacie Memorial Fellowships (A.D.), the Alexander von Humboldt Fellowship (A.D.), the Canada Research Chairs Program (A.D.), NSERC, Canada Foundation for Innovation (CFI), and the CIFAR Quantum Materials Program. Use of the Advanced Photon Source at Argonne National Laboratory was supported by the DOE, Office of Science, Office of Basic Energy Sciences, under contract no.~DE-AC02-06CH11357. This research also used resources of the Advanced Light Source, which is a DOE Office of Science User Facility under contract no.~DE-AC02-05CH11231.

\bibliography{TiSe2Refs}

\end{document}